\documentclass[aps,prl,twocolumn,showpacs,superscriptaddress]{revtex4-2}

\usepackage{amsmath,amssymb}
\usepackage{graphicx}
\usepackage{hyperref}
\usepackage{xcolor}
\DeclareMathOperator{\tr}{tr}
\begin{document}

\title{Adaptive Punishment in Social Dilemmas}

\author{Xingfu Ke}
\affiliation{Alibaba Research Center for Complexity Sciences, Hangzhou Normal University, Hangzhou 311121, Zhejiang, China}
\author{Hao Yu}
\affiliation{Alibaba Research Center for Complexity Sciences, Hangzhou Normal University, Hangzhou 311121, Zhejiang, China}
\author{Xiao-Pu Han}
\affiliation{Alibaba Research Center for Complexity Sciences, Hangzhou Normal University, Hangzhou 311121, Zhejiang, China}
\author{Yi-Cheng Zhang}
\affiliation{International Research Center for Complexity Sciences, Hangzhou International Innovation Institute, Beihang University, Hangzhou 311115, Zhejiang, China}
\author{Fanyuan Meng}
\email{fanyuan.meng@hotmail.com}
\affiliation{Alibaba Research Center for Complexity Sciences, Hangzhou Normal University, Hangzhou 311121, Zhejiang, China}

\date{\today}

\begin{abstract}
We introduce a coevolutionary framework in which punishment intensity dynamically adapts to the fraction of cooperators in the population. Unlike static models, adaptive punishment reshapes the effective payoff landscape, driving transitions among canonical games, including the Prisoner’s Dilemma, Harmony, Stag Hunt, and Chicken games. Analytical results reveal rich dynamical behaviors such as coexistence, bistability, limit cycle and Hopf bifurcation. These findings highlight adaptive punishment as a robust mechanism for sustaining cooperation by the coevolutionary feedback and offer insights into institutional design, ecological interactions, and social governance.
\end{abstract}

\maketitle

\section{Introduction}
The emergence and persistence of cooperation in social dilemmas remain a central puzzle across biology, economics, social sciences and physics \cite{nowak2012evolving,schuster2004cooperate,fehr2007human,perc2017statistical}. Classic studies, from Hardin’s “tragedy of the commons” to Ostrom’s institutional theory, highlight how individual rationality undermines collective welfare~\cite{feeny1990tragedy,Ostrom1990}. Evolutionary game theory provided a quantitative framework to explain cooperative behavior through mechanisms such as kin selection, reciprocity, network reciprocity, and multilevel selection~\cite{Hamilton1964,MaynardSmith1982,Axelrod1984,Nowak2006,Lehmann2007}. Despite these advances, maintaining cooperation in populations of self-interested individuals remains challenging when temptation to defect is strong and external regulation is limited.

Punishment and reward have long been recognized as effective mechanisms to sustain cooperation~\cite{Fehr2002,FehrFischbacher2004,Henrich2006,Fowler2005}. Laboratory and field experiments show that altruistic and third-party punishment, reputation, and indirect reciprocity can stabilize collective action~\cite{Milinski2002,NowakSigmund1998,Panchanathan2004}. Theoretical models formalized how incentives modify payoffs and alter evolutionary trajectories~\cite{Sigmund2001,Boyd2003,Gintis2000,NowakMay1992}. However, most frameworks treat sanctioning as exogenous and static, entering as a fixed cost or benefit independent of behavioral context, while real institutions display adaptive feedback: enforcement often intensifies when defection spreads and relaxes when cooperation prevails~\cite{Perc2010,Helbing2010,Szolnoki2011}. This missing feedback loop obscures how enforcement and cooperation co-shape social dynamics.

Here we propose a minimal coevolutionary model in which enforcement intensity evolves dynamically with the population’s cooperation level. In contrast to previous coevolutionary approaches that modified network topology or partner selection~\cite{Pacheco2006,Santos2005,Perc2010}, our formulation introduces a continuously varying enforcement variable that alters the effective payoff matrix, driving endogenous transitions among canonical two-player games (Harmony, Stag Hunt, Prisoner’s Dilemma, Chicken)~\cite{Rapoport1965,Nowak2006book}. Analytical and numerical analyses reveal bistability between full cooperation and full defection, interior fixed points sustaining partial cooperation, and Hopf bifurcations that generate oscillatory cycles of cooperation and enforcement. The system’s behavior is governed by the net effect of sanctioning, balancing punishment and reward, which determines whether adaptive feedback amplifies or suppresses cooperation.

Our framework advances the theory of cooperative dynamics in three ways. First, it formalizes the self-organization of institutional enforcement as an endogenous process rather than an external control. Second, it uncovers a universal route by which adaptive regulation reshapes the payoff structure, allowing populations to traverse distinct game classes within a single dynamical system. Third, it establishes a minimal and analytically transparent model connecting contingent sanctioning observed in experiments~\cite{Fehr2002,Henrich2006} with broader theories of coevolutionary games and adaptive networks~\cite{Perc2010,Szabo2007,Gomez2011,PercSzolnoki2010}. Together these results reveal how adaptive punishment and reward can stabilize cooperation through self-organized feedback loops, offering testable predictions for institutional design in social, ecological, and microbial systems.

\section{The Model} We consider a well-mixed population of players engaging in repeated pairwise interactions, each adopting either cooperation (C) or defection (D). The payoff matrix is
\begin{equation}
\begin{array}{c|cc}
 & C & D \\
\hline
C & 1 & S + \eta\,\beta\,T \\
D & (1-\beta)T & 0
\end{array},
\label{eq:payoff}
\end{equation}
where $T$ and $S$ denote the classical temptation and sucker payoffs. The parameter $\beta\in[0,1]$ is the punishment intensity, the effective strength of institutional sanctions applied in each interaction. The tunable parameter $(\beta \in [0,1])$ denotes the strength of institutional sanctioning, whereas
$\eta \in (-\infty, 1]$ specifies its alignment, the extent to which enforcement correctly targets defectors. Positive $(\eta > 0)$ corresponds to \emph{aligned sanctioning}, where defectors are penalized and cooperators receive compensatory benefits. The neutral case $(\eta = 0)$ represents undirected enforcement with no behavioral discrimination. Negative $(\eta < 0)$ captures \emph{misaligned sanctioning}, including antisocial punishment and group-level penalties in which cooperators also incur losses. Thus a single parameter $\eta$ continuously spans reward-based, punishment-based, and collectively imposed sanctions, allowing enforcement accuracy to coevolve with cooperation within one unified dynamical framework.

Let $x$ be the fraction of cooperators. The payoff difference between cooperators and defectors is then
\begin{equation}
\pi_C-\pi_D = x\big[1-(1-\beta)T\big] + (1-x)(S+\eta\beta T).
\label{eq:pi-diff}
\end{equation}

Strategy evolution follows the replicator dynamics
\begin{equation}
\dot{x}=x(1-x)(\pi_C-\pi_D),
\label{eq:Dx}
\end{equation}
while the punishment intensity coevolves according to
\begin{equation}
\dot{\beta}=\beta(1-\beta)\big(1-x-\theta x\big),
\label{eq:Dbeta}
\end{equation}
where $\theta > 0$ quantifies the relative strength of cooperators in dampening 1 versus that of defectors in amplifying it.


\paragraph{Stability Analysis} The steady states of the system correspond to the fixed points of Eqs.~\eqref{eq:Dx}–\eqref{eq:Dbeta}.
Linear stability is determined by the Jacobian matrix $J(x, \beta)$:
\begin{equation}
\begin{pmatrix}
(1\!-\!2x)(\pi_C\!-\!\pi_D) + x(1\!-\!x)K(\beta) & x(1\!-\!x)\, T\left(\eta\!+\!x(1\!-\!\eta)\right) \\
-\beta(1\!-\!\beta)(1\!+\!\theta) & (1\!-\!2\beta)\left((1\!-\!x)\!-\!\theta x\right)
\end{pmatrix}
\label{eq:Jacobian}
\end{equation}
with $K(\beta) = 1-T-S + (1-\eta)T \beta$.

The four corner points are $(0,0)$, $(0,1)$, $(1,0)$, and $(1,1)$. Among them, $(0,0)$ and $(1,1)$ are always unstable. 

The point $(0,1)$ is stable when 
\begin{equation}
    S+ \eta T < 0,
    \label{eq:cond_(0,1)}
\end{equation}
while $(1,0)$ is stable when 
\begin{equation}
    T < 1. 
    \label{eq:cond_(1,0)}
\end{equation}


A boundary fixed point appears at $(x^L,0)$ with
$x^L=S/(T+S-1)$, satisfying $\pi_C - \pi_D=0$ at $\beta=0$. The trace and determinant of the Jacobian at this point are
\begin{equation}
    \begin{cases}
        \det J^L=x^L(1-x^L)(1-T-S)(1-x^L-\theta x^L),\\
\tr J^L=x^L(1-x^L)(1-T-S) + (1-x^L - \theta x^L).
    \end{cases}
\end{equation}

Thus, $(x^L,0)$ exists and is stable when $0<x^L<1$, $\tr J^L<0$ and $\det J^L>0$, yielding 
\begin{equation}
    S>0, \quad 1< T<1+S \theta.
    \label{eq:cond_(xL,0)}
\end{equation}




An interior fixed point $(x^*, \beta^*)$ exists when $1-x^* - \theta x^*=0$ and $\pi_C - \pi_D=0$, yielding
\begin{equation}
x^*=\frac{1}{1+\theta}, \quad 
\beta^*= \frac{T-1-S\theta}{T(1+\eta \theta)}.
\end{equation}

The corresponding trace and determinant are
\begin{equation}
    \begin{cases}
        \det J^*=x^*(1-x^*)\beta^*(1-\beta^*)T(\eta+x^*(1-\eta)),\\
\tr J^*=(1-x^*)K(\beta^*).
    \end{cases}
\end{equation}

Thus, the internal fixed point $(x^*, \beta^*)$ is stable if and only if
\begin{equation}
        \eta > \frac{S}{1-T}, \quad T>1, \quad T>1+ S \theta.
    \label{eq:cond_inner}
\end{equation}




\begin{figure}[h]
\includegraphics[width=\linewidth]{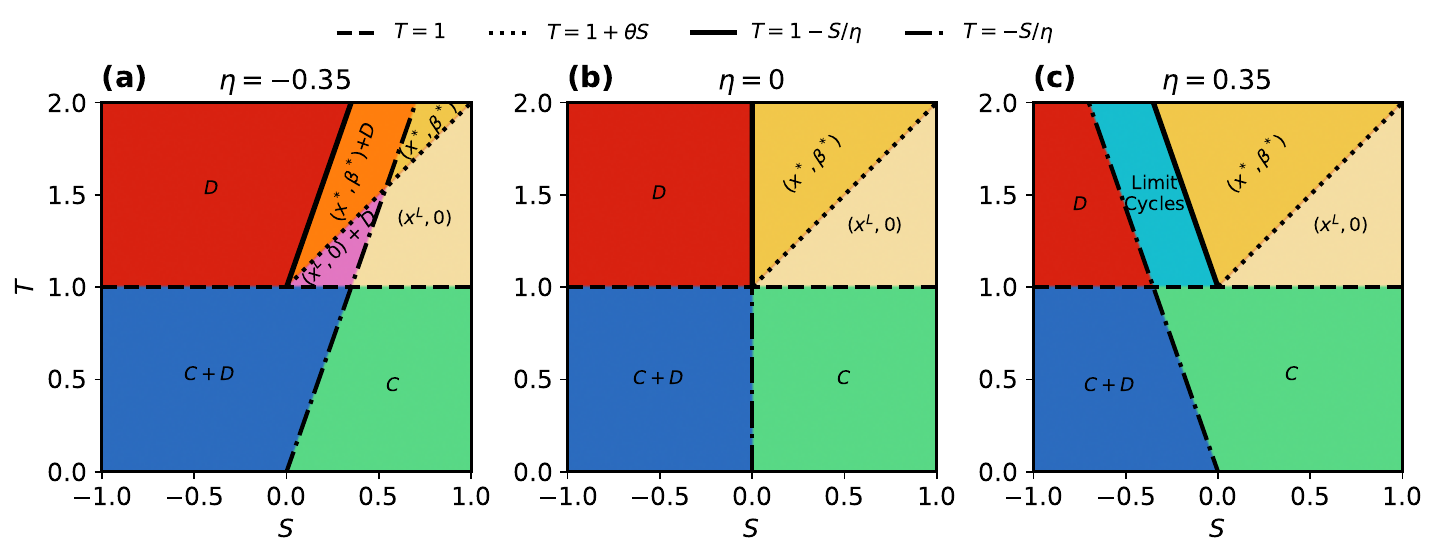}
\caption{\label{fig:1} 
Phase diagram in the $(T,S)$ plane. (a-c) $\eta = \{-0.35,0,0.35\}$.}
\end{figure}



\section{Dynamical Regimes}
\begin{figure}[h]
\includegraphics[width=\linewidth]{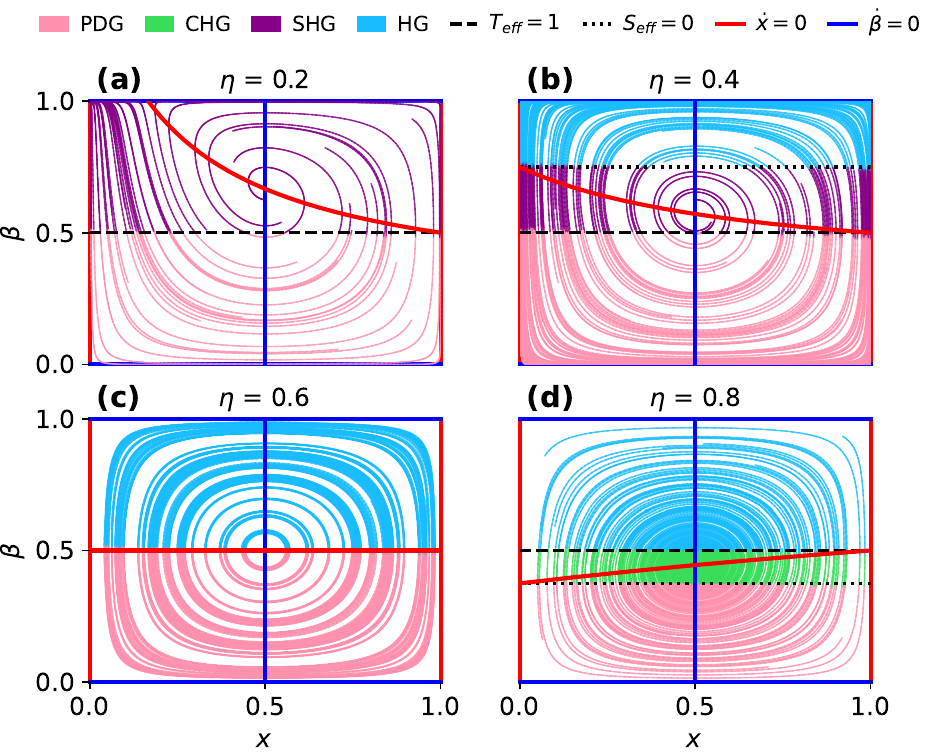}
\caption{\label{fig:4} 
Phase plane for an initial PDG ($T=2$, $S=-0.6$) with $\theta=1$.  
(a-d) $\eta=\{0.2,0.4,0.6,0.8\}$.}
\end{figure}

\begin{figure}[h]
\includegraphics[width=\linewidth]{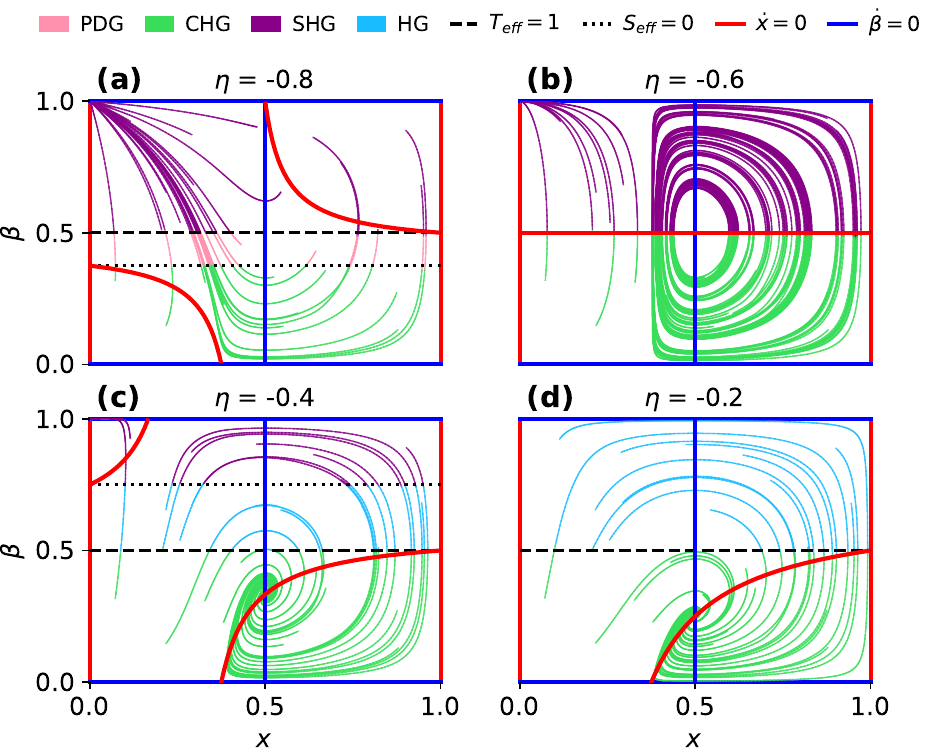}
\caption{\label{fig:5} 
Phase plane for an initial CHG ($T=2$, $S=0.6$) with $\theta=1$. (a-d) $\eta=\{-0.8,-0.6,-0.4,-0.2\}$.}
\end{figure}

Fig.~\ref{fig:1} illustrates how sanction alignment, controlled by $\eta$, restructures the dynamical landscape in the $(T,S)$ plane. For misaligned or collective sanctioning ($\eta<0$, Fig.~\ref{fig:1}(a)), Eqs.~\ref{eq:cond_(0,1)}–\ref{eq:cond_(1,0)} carve the plane into distinct regimes in which enforcement either stabilizes or destabilizes cooperative behavior. When $T<1$ and $T>-S/\eta$, the system exhibits bistability between full cooperation ($C$) and full defection ($D$), whereas for $T<1$ and $T\le -S/\eta$, only $C$ remains stable. For $T>1$, misalignment induces a more intricate structure: if $T>1-S/\eta$, defection dominates; if $T<1-S/\eta$ and $T>-S/\eta$, an interior fixed point $(x^*,\beta^*)$ coexists with $D$; if $T<1+S\theta$ and $T>-S/\eta$, the boundary equilibrium $(x^L,0)$ coexists with $D$; and along $T=1+S\theta$, a marginal equilibrium $(1/(1+\theta),0)$ appears. In the region $T>1$ and $T\le -S/\eta$, the system transitions between $(x^*,\beta^*)$ and $(x^L,0)$ depending on whether $T$ lies above or below $1+S\theta$. Moreover, at the critical line $T=1-S/\eta$, a Hopf bifurcation emerges, generating a continuum of closed orbits (Fig.~\ref{fig:5}(b)) that reflect oscillatory feedback between cooperation and enforcement. When sanctioning becomes undirected ($\eta=0$, Fig.~\ref{fig:1}(b)), for $T>1$, the two lines $T=1-S/\eta$ and $T=-S/\eta$ overlap and become vertical, letting two bistable regions of $(x^*,\beta^*) + D$ and $(x^L,0) + D$ vanish. In contrast, aligned prosocial sanctioning ($\eta>0$, Fig.~\ref{fig:1}(c)), these two bistable regions are replaced by limit cycles: for $T>1$, $T<1-S/\eta$, and $T>-S/\eta$, cooperation and enforcement jointly oscillate through self-organized feedback, and along $T=1-S/\eta$ a Hopf bifurcation again generates families of closed orbits (Fig.~\ref{fig:4}(c)).

\section{Game Transitions}
We next demonstrate how the coevolving \(\beta\)-\(x\) feedback shapes the system’s dynamical trajectories, while reshaping the game-class boundaries characterized by the critical surfaces \(T_{\mathrm{eff}} = 1\) and \(S_{\mathrm{eff}} = 0\), which correspond to \(\beta_T = 1 - 1/T\) and \(\beta_S = -S/(\eta T)\), across the four canonical \((2 \times 2)\) games. 

\subsection{Prisoner’s Dilemma Game (PDG): Baseline for Cross-Game Cycling}
PDG is defined by \(T_{\text{eff}}>1\) and \(S_{\text{eff}}<0\), a regime where defection is individually optimal but collectively costly. We fix \(T=2\) and \(S=-0.6\) to initialize the system in PDG for small \(\beta\) and moderate \(\eta\), establishing a baseline for tracking adaptive transitions. For weak aligned sanctioning (\(\eta=0.2\)), the stability of the \((0,1)\) fixed point (full defection with maximum punishment) is preserved (\(S+\eta T=-0.2<0\)), and transitions are restricted to PDG (for \(\beta<0.5\), i.e., \(T_{\text{eff}}>1\)) and Stag Hunt Game (SHG, for \(\beta>0.5\), \(T_{\text{eff}}<1\)). Strengthening alignment (\(\eta=0.4\)) eliminates \((0,1)\) stability (\(S+\eta T=0.2>0\)), enabling multi-game cycling: PDG (below \(\beta_T=0.5\)), SHG (between \(\beta_T=0.5\) and \(\beta_S=0.75\), where \(S_{\text{eff}}>0\)), and Harmony Game (HG, above \(\beta_S=0.75\)). At the Hopf bifurcation threshold (\(\eta=0.6\), satisfying \(T=1-S/\eta\)), \(\beta_T=\beta_S=0.5\), simplifying transitions to PDG-HG toggling via neutrally stable closed orbits. For strong alignment (\(\eta=0.8\)), the system converges to a stable interior fixed point (\(x^*=1/2\), \(\beta^*=4/9\)) in the Chicken Game (CHG) regime (\(T_{\text{eff}}>1\), \(S_{\text{eff}}>0\)), demonstrating PDG’s role as a launchpad for all three other game classes.  

\begin{figure}[h]
\includegraphics[width=\linewidth]{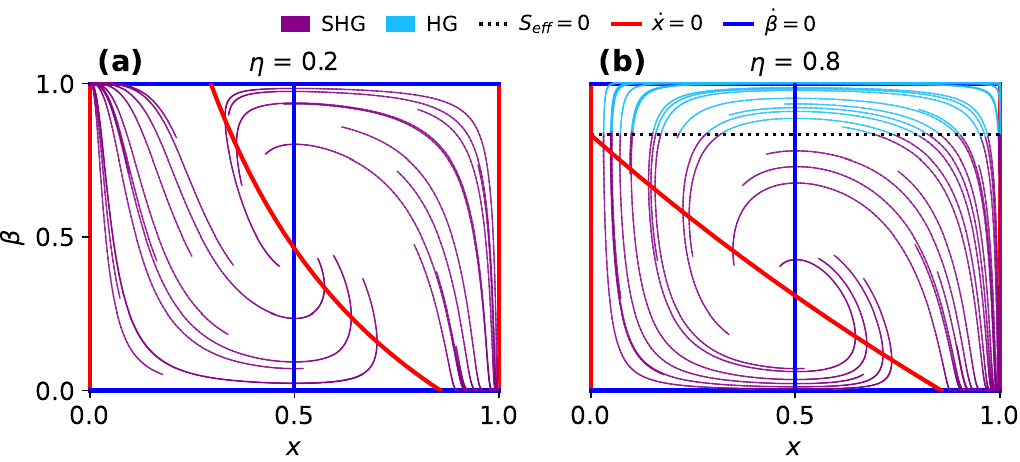}
\caption{\label{fig:3} 
Phase plane for an initial SHG ($T=0.9$, $S=-0.6$) with $\theta=1$. (a) $\eta=0.2$; (b) $\eta=0.8$.}
\end{figure}

\subsection{Stag Hunt Game (SHG): Cooperative Stabilization via Threshold Crossing}
SHG is characterized by \(0<T_{\text{eff}}<1\) and \(S_{\text{eff}}<0\), balancing risk (sucker payoff) and reward (cooperative gain). We fix \(T=0.9\) (ensuring \(T_{\text{eff}}<1\) for all \(\beta\)) and \(S=-0.6\) to initialize SHG, mirroring PDG’s sucker payoff structure but weakening temptation. For weak alignment (\(\eta=0.2\)), both \((0,1)\) (full defection) and \((1,0)\) (full cooperation) are stable, and SHG persists across all \(\beta\) (since \(S_{\text{eff}}<0\) holds). This contrasts with PDG’s cycling, as SHG’s muted temptation (\(T=0.9<1\)) restricts transitions. Strengthening alignment (\(\eta=0.8\)) destabilizes \((0,1)\) (\(S+\eta T=0.12>0\)), triggering conditional transitions: if \(\beta\) exceeds \(\beta_S=5/6\) (where \(S_{\text{eff}}>0\)), the trajectory shifts to HG before reverting to SHG and converging to \((1,0)\); otherwise, it remains in SHG while approaching full cooperation. This highlights SHG as a "stepping stone" between PDG’s defection dominance and HG’s cooperative optimality, sharing PDG’s \(\beta_S\) threshold but with reduced sensitivity to \(\beta\) due to weaker \(T\).

\subsection{Chicken Game (CHG): Reverse Transitions Under Misaligned Sanctioning}
CHG is defined by \(T_{\text{eff}}>1\) and \(S_{\text{eff}}>0\), a regime where mutual defection is costly but unilaterally defecting yields temptation. We fix \(T=2\) and \(S=0.6\) to initialize CHG, reversing PDG/SHG’s sucker payoff sign to explore misaligned sanctioning (\(\eta<0\)). For strong misalignment (\(\eta=-0.8\)), \((0,1)\) is stabilized (\(S+\eta T=-1<0\)), driving transitions from CHG (below \(\beta_S=3/8\), \(S_{\text{eff}}>0\)) to PDG (between \(\beta_S=3/8\) and \(\beta_T=1/2\), \(S_{\text{eff}}<0\)) and then to SHG (above \(\beta_T=1/2\), \(T_{\text{eff}}<1\))—a reverse of PDG’s aligned transitions. Weakening misalignment (\(\eta=-0.6\)) aligns \(\beta_S=\beta_T=0.5\), producing Hopf bifurcation-driven CHG-SHG toggling. Further reducing misalignment (\(\eta=-0.4\)) introduces bistability between \((0,1)\) and a stable interior CHG fixed point (\(x^*=1/2\), \(\beta^*=1/3\)), while the weakest misalignment (\(\eta=-0.2\)) eliminates \((0,1)\) and converges to CHG. CHG thus mirrors PDG’s dynamics but under misaligned sanctions, confirming that \(\eta\) (sanction alignment) dictates the direction of game transitions.

\begin{figure}[h]
\includegraphics[width=\linewidth]{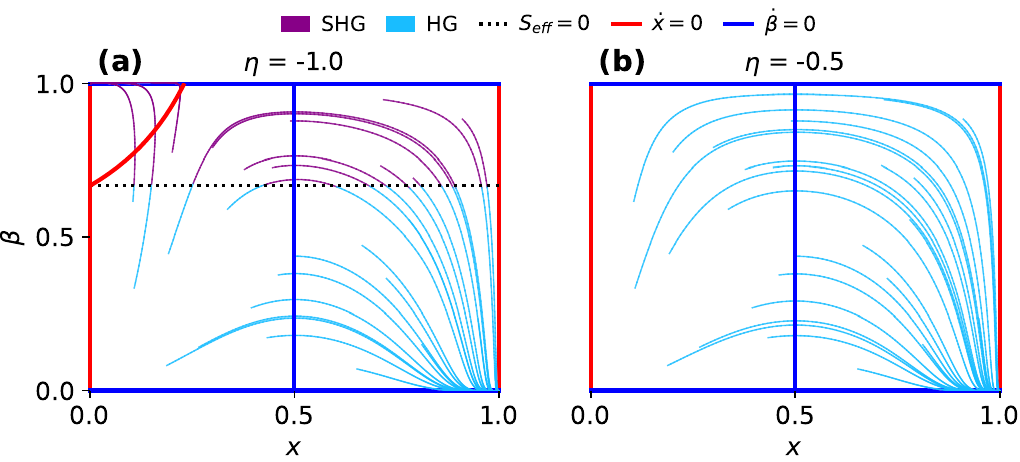}
\caption{\label{fig:2} 
Phase plane for an initial HG ($T=0.9$, $S=0.6$) with $\theta=1$. (a) $\eta=-1$; (b) $\eta=-0.5$.}
\end{figure}

\subsection{Harmony Game (HG): Cooperative Terminus of Adaptive Trajectories}
HG is the cooperative optimum, defined by \(0<T_{\text{eff}}<1\) and \(S_{\text{eff}}>0\), where cooperation is both individually and collectively beneficial. We fix \(T=0.9\) and \(S=0.6\) to initialize HG, combining SHG’s weak temptation and CHG’s positive sucker payoff. For strong misaligned sanctioning (\(\eta=-1\)), \((0,1)\) is stabilized (\(S+\eta T=-0.3<0\)), triggering transitions to SHG (above \(\beta_S=2/3\), \(S_{\text{eff}}<0\)) before reverting to HG and converging to \((1,0)\)—echoing SHG’s conditional transitions. Weakening misalignment (\(\eta=-0.5\)) destabilizes \((0,1)\) (\(S+\eta T=0.15>0\)), and the system remains in HG throughout its trajectory to \((1,0)\). HG thus serves as the terminal regime for adaptive transitions from PDG, SHG, and CHG: aligned sanctioning (PDG/SHG) and reduced misalignment (CHG) both drive convergence to HG, while strong misalignment temporarily pushes HG to SHG (mirroring CHG’s PDG/SHG transitions). This unifies all four games under a common adaptive logic: enforcement alignment (\(\eta\)) and intensity (\(\beta\)) steer trajectories toward HG’s cooperative equilibrium.

\subsection{Discussion and Conclusion} We have developed a coevolutionary framework in which the punishment intensity $(\beta \in [0,1])$ dynamically adapts to the cooperation fraction $(x)$, while the alignment parameter $(\eta \in (-\infty, 1])$ continuously spans aligned prosocial $(\eta > 0)$, neutral $(\eta = 0)$, and misaligned or collective $(\eta < 0)$ sanctioning. This bidirectional feedback reshapes the effective payoffs $T_{\rm eff}$ and $S_{\rm eff}$, enabling endogenous transitions among the four canonical games: SHG, HG, PDG, and CHG. Cross-game boundaries are governed jointly by $S_{\rm eff} = 0$ and $T_{\rm eff} = 1$, and adaptive punishment produces rich dynamics absent in static models, including bistability, Hopf bifurcations, limit cycles, and stable interior fixed points $(x^*, \beta^*)$ that sustain cooperation. Beyond theory, our results provide a quantitative basis for adaptive enforcement: tuning $\eta$ can stabilize cooperation in social, ecological, or institutional systems, while the coevolutionary $\beta$–$x$ feedback enables self-regulation without rigid sanctions, offering a robust mechanism for sustaining cooperation in complex populations.

\begin{acknowledgments}
This work is supported by the National Natural Science Foundation of China (Grant Nos.12505044 and 52374013). We thank Prof. Zhigang Zheng for helpful discussions. During the preparation of this work the author(s) used GPT in order to improve language of the manuscript. After using these tools, the authors reviewed and edited the content as needed and take full responsibility for the content of the publication.
\end{acknowledgments}
\bibliographystyle{apsrev4-2}
\bibliography{prl.bib}

\end{document}